\documentclass[conference]{IEEEtran}
\IEEEoverridecommandlockouts

\usepackage{cite}
\usepackage{amsmath,amssymb,amsfonts}
\usepackage{algorithmic}
\usepackage{graphicx}
\usepackage{textcomp}
\usepackage{xcolor}
\def\BibTeX{{\rm B\kern-.05em{\sc i\kern-.025em b}\kern-.08em
    T\kern-.1667em\lower.7ex\hbox{E}\kern-.125emX}}

\usepackage{latexsym}
\usepackage{hyperref}
\usepackage{gensymb}
\usepackage{siunitx}
\usepackage{tikz}
\usepackage{etoolbox}
\usepackage{enumitem}
\usepackage{esvect}
\usepackage{mathtools}
\usetikzlibrary{chains,arrows,arrows.meta,calc,positioning}
\usepackage{multirow}
\usepackage{comment}
\usepackage[normalem]{ulem}
\usepackage{upquote}
\usepackage{caption}
\usepackage{subcaption}
\usepackage{tikz}
\usetikzlibrary{angles,quotes,positioning,calc,shapes.geometric, arrows.meta, shadows.blur}
\usepackage{contour}            
\contourlength{0.7pt}           

\usepackage{cleveref}
\usepackage{bbm}
\usepackage{booktabs}
\usepackage{tabularx}
\usepackage{dblfloatfix}

\DeclareMathOperator*{\argmax}{arg\,max}

\def\Exp{\mathbb{E}\,}

\newcommand{\subsf}{\sf \scriptscriptstyle}
\newcommand{\bs}{\boldsymbol}
\newcommand{\mc}{\mathcal}
\newcommand{\wh}{\widehat}
\newcommand{\ov}{\overline}

\newcommand{\herm}{^{\text{\sf H}}}

\newcommand{\C}{\mathbb{C}}
\newcommand{\R}{\mathbb{R}}

\begin{document}

\title{Beyond Point Estimates: Likelihood-Based Full-Posterior Wireless Localization
}

\author{
\IEEEauthorblockN{Haozhe Lei$^{*1}$, Hao Guo$^{1,2}$, Tommy Svensson$^{2}$, and Sundeep Rangan$^{1}$}
\IEEEauthorblockA{$^1$NYU WIRELESS, Tandon School of Engineering, New York University, Brooklyn, NY 11201, USA\\
$^2$Department of Electrical Engineering, Chalmers University of Technology, Gothenburg, Sweden\\
\{hl4155, hg2891, srangan\}@nyu.edu; tommy.svensson@chalmers.se}
}

\maketitle

\begingroup
\renewcommand\thefootnote{}
\footnotetext{$^{\ast}$ Corresponding author: Haozhe Lei (hl4155@nyu.edu).}
\endgroup

\begin{abstract}
Modern wireless systems require not only position estimates, but also quantified uncertainty to support planning, control, and radio resource management. We formulate localization as posterior inference of an unknown transmitter location from receiver measurements.
We propose \emph{Monte Carlo Candidate-Likelihood Estimation (MC-CLE)}, which trains a neural scoring network using Monte Carlo sampling to compare true and candidate transmitter locations.
We show that 
in line-of-sight simulations with a multi-antenna receiver, MC-CLE learns critical 
properties including angular ambiguity and front-to-back antenna patterns.  MC-CLE also achieves lower cross-entropy loss relative to a uniform baseline and Gaussian posteriors.  alternatives under a uniform-loss metric.
\end{abstract}


\section{Introduction}

Wireless localization plays a central role in a wide range of applications and is now an integral component of the 5G standard~\cite{TR38.859,Rel18Enhance24}. This paper considers the problem of estimating the location $\bs{x}^t$ of a wireless transmitter using measurements collected by one or more receivers. Conventional approaches typically yield a point estimate $\wh{\bs{x}}^t$ based on some observations $\bs{y}$. However, in many practical scenarios, these estimates are accompanied by significant uncertainty, and downstream tasks increasingly demand a richer characterization of that uncertainty,
or so-called uncertainty awareness
\cite{wang2025uncertainty}.
Applications such as robust path planning, risk-aware control, and adaptive beamforming or handover decisions~\cite{zhang2024rloc,shamsfakhr2022indoor,LeiICRA2024,Lei2025DTWIN,Li25RLPhysics,Shahmansoori18} rely not just on a single location guess, but on an understanding of where the transmitter could plausibly be located—and with what confidence.

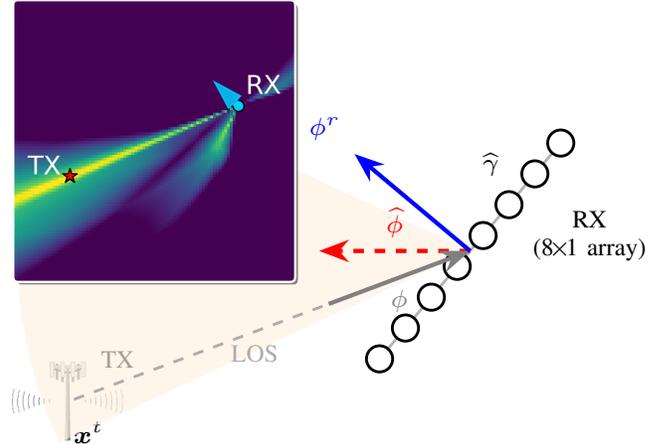
\begin{figure}[t]
  \centering
  \resizebox{\linewidth}{!}{%
  \begin{tikzpicture}[
      >=Stealth,
      every node/.style={font=\scriptsize}
    ]
    \tikzset{
      elem/.style={fill=white, draw=black, line width=0.8pt}
    }

    \def\arrayAngle{50}     
    \def\nElements{8}       
    \def\elemSpacing{0.40}  
    \def\elemRadius{0.13}   
    \pgfmathsetmacro{\halfLen}{(\nElements-1)/2*\elemSpacing}

    \coordinate (TX) at (0,0);
    \node[inner sep=0pt, anchor=center] (TXicon) at (TX)
      {\includegraphics[width=15mm]{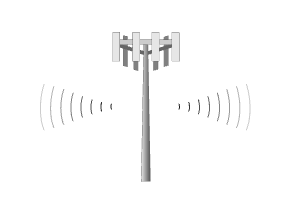}};
    \node[right=6pt, yshift=12pt] at (TX) {TX};
    \node[below=2pt, xshift=6pt] at (TX) {$\bs{x}^t$};

    \coordinate (RXc) at (4,1.5);

    \path[fill=orange!12,opacity=0.65]
      (RXc) -- ++(180-25:4.5) arc (155:205:4.5) -- cycle;

    \draw[gray!60, line width=0.8pt]
      ($(RXc)+(\arrayAngle:{-\halfLen})$) --
      ($(RXc)+(\arrayAngle:{\halfLen})$);
    \foreach \k in {0,...,\numexpr\nElements-1\relax}{
      \pgfmathsetmacro{\off}{(\k-(\nElements-1)/2)*\elemSpacing}
      \path ($(RXc)+(\arrayAngle:\off)$) coordinate (E\k);
      \filldraw[elem] (E\k) circle (\elemRadius);
    }

    \draw[dashed, gray!65, thick]
      ($(RXc)!.3cm!(TX)$) -- (TX) node[midway, below] {LOS};

    \draw[->,very thick,red,dashed]
      (RXc) -- ++(180:1.5cm) node[midway, above] {$\widehat{\phi}$};  
    \draw[->,gray,very thick]
      ($(RXc)!1.5cm!(TX)$) -- (RXc)
      node[midway, below=-0.5pt] {$\phi$};                            
    \draw[->,very thick,blue]
      (RXc) -- ++(140:1.5cm) node[above left] {$\phi^r$};             

    \node[anchor=west] at ($(RXc)+(0.00,0.85)$) {\contour{white}{$\widehat{\gamma}$}};
    \node[align=center, anchor=west]
      at ($(RXc)+(0.50,0.15)$) {RX\\(8$\!\times\!$1 array)};

    \node[anchor=north east] at ($(RXc)+(-15mm,27mm)$)
      {\includegraphics[width=3.0cm]{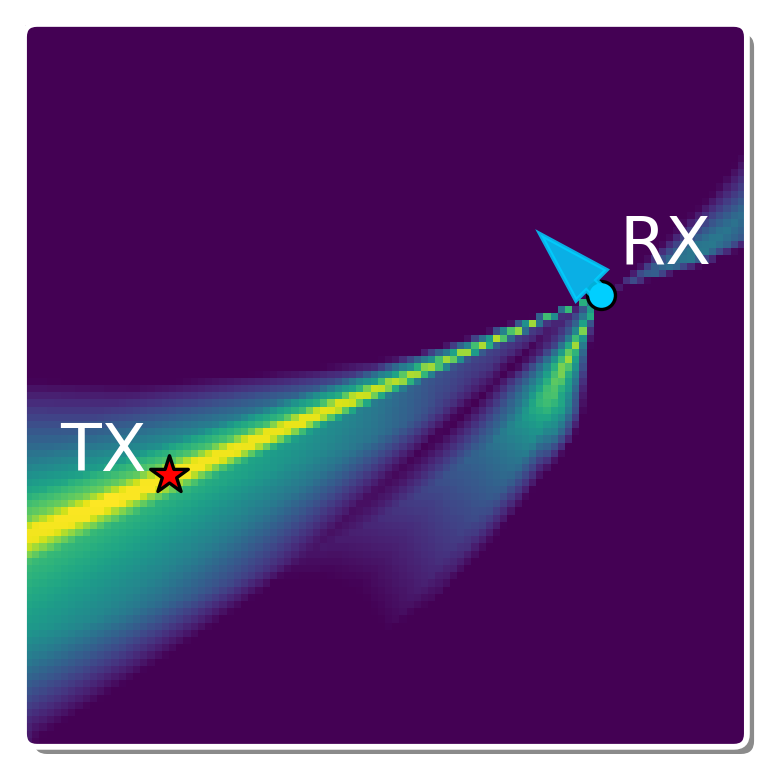}};
  \end{tikzpicture}%
  }
  \caption{Example learned posterior $p(\bs{x}^t\!\mid\!\bs{y})$ for an unknown transmitter location $\bs{x}^t$ from receiver measurements $\bs{y}$ using an $8{\times}1$ uniform linear array. The posterior is visualized as a heat map.}
  \label{fig:loc_geometry}
\end{figure}

To meet these demands, this work considers posterior inference instead of point estimation.   Specifically, we aim to estimate the posterior distribution 
\begin{equation} \label{eq:posterior} 
    p(\bs{x}^t|\bs{y}), 
\end{equation} 
which represents the probability of the transmitter being at location $\bs{x}^t$ given the observations $\bs{y}$. For given observations $\bs{y}$, the posterior $p(\bs{x}^t|\bs{y})$ can be visualized as a spatial heat map, revealing regions of high and low likelihood across the search space. \Cref{fig:loc_geometry} illustrates such a heat map generated by our proposed method, Monte Carlo Conditional Likelihood Estimation (MC-CLE), which we introduce in detail below.


\paragraph*{Challenges, even under line-of-sight} We consider the problem of estimating the posterior distribution \eqref{eq:posterior} from an unknown transmitter location from receiver measurements in a line-of-sight (LOS) channel. The receiver is equipped with an array and performs matched filtering to detect the angle and SNR of a known transmitted signal, and the posterior is to be estimated from these measurements along with the receiver pose.  Even in this simple LOS case, we show that the posterior distribution can be complex: 
(1) Most importantly, the posterior is frequently a mixture distribution due to false alarms—typically manifesting as a dominant peak corresponding to the true path and a secondary residual distribution arising from noise-induced detections. (2) Array geometries often introduce angular ambiguities; for instance, uniform linear arrays (ULAs) produce identical responses at angles $\phi$ and $\pi - \phi$. (3) Antenna elements, such as patch antennas, exhibit strong directionality, making the received signal highly sensitive to the receiver’s orientation.

\subsection*{Contributions}
Our contributions are as follows:
(i) We formalize localization as posterior density estimation over a 2D scene (\Cref{sec:prob_form}) and discuss several challenges for this problem even in the LOS case (\Cref{sec:challenges}). 
(ii) We introduce \emph{Monte Carlo Candidate-Likelihood Estimation (MC-CLE)}, which learns the likelihood by comparing the score of the true location to randomly selected candidate locations.
(iii) MC-CLE is evaluated with realistic antenna models, including antenna patterns with significant front-to-back gain.  It is shown that MC-CLE captures many of the key features of the conditional distribution, including distribution mixtures, multi-modalities from angular ambiguity, and directionality.  The method outperforms parametric methods such as Gaussian conditional distributions (\Cref{sec:experiments}).

\subsection*{Prior work}  
The overwhelming majority of conventional RF localization methods provide point estimates.  The characterization of the uncertainty of these estimates is much less understood.  The Cram{\'e}r-Rao bound
is widely-used for bounding the error
of unbiased estimates 
\cite{godrich2008cramer}.  However, applying the CR bound requires a model of the measurements and is typically only achieved in the high SNR regime.  Related  work
\cite{zhang2024rloc} learns the standard deviation of angular errors;
\cite{shamsfakhr2022indoor} quantifies errors in triangulization; and
\cite{muppirisetty2015spatial}
characterizes the effect of location uncertainty in the anchor nodes.
None of these methods provide a full spatial characterization of the spatial posterior as considered in this work.  

Bayesian approaches maintain posterior maps and fuse heterogeneous measurements via particle filtering~\cite{Arulampalam02} and factor-graph belief propagation~\cite{Wymeersch09}, with von~Mises–Fisher angle noise improving likelihood shape~\cite{Nurminen18}, yet they rely on well-specified likelihoods, fixed grids or tuned proposals, and significant computation, and can be biased under model mismatch.

The general problem of learning conditional distributions is a classic problem in statistics.   Non-parametric techniques, such as kernel estimation, are widely-used -- see the survey \cite{silverman2018density}.
Closely related are so-called localization methods \cite{hall1999methods} that locally fit logistic models using kernel weights.
As non-parametric methods generally suffer from the curse of dimensionality, this work attempts to learn a parametric approximation of the likelihood $g_\theta(\bs{x}^t,\bs{y}) \approx \ln p(\bs{y}|\bs{x}^t)$.  The well-known challenge in the parametric method is that the normalization constant or partition function $Z_\theta(\bs{y})$ in \eqref{eq:Zy} involves an integral that does not have a closed-form expression.   One approach is to discretize the integral
\cite{lindsey1974comparison,gao2022lincde}.  As we discuss below, our method can be seen as Monte-Carlo sampling.



\section{Problem Formulation}
\label{sec:prob_form}
\subsection{Posterior Density Estimation }
Consider the problem of locating a transmitter (TX) at an unknown position, $\bs{x}^t \in \R^d$.  For simplicity, we consider the 2D localization problem shown in \Cref{fig:loc_geometry}, so $d=2$.  We assume that TX localization has a known prior $p_0(\bs{x}^t)$.  Typically, we take $p_0(\bs{x}^t)$ to be uniform in some compact set $\mc{A} \subset \R^d$, but any prior distribution may be used.  A receiver (RX) is at a known 2D \emph{pose}, $(\bs{x}^r,\phi^r)$, where $\bs{x}^r \in \R^d$ is the RX position and $\phi^r$ is its azimuth orientation.

To locate the TX, the TX broadcasts some periodic signals.  For example, the TX could broadcast IEEE 802.11 Wi-Fi beacon frames every 100 ms~\cite{IEEE80211-2020} in a small room, IEEE 802.15.4z UWB blink pulses in large indoor warehouses~\cite{IEEE802154z-2020}, or GPS L1 C/A codes every 1 ms~\cite{GPSsignals} or 5G NR SSBs every 20 ms~\cite{3GPP38211} in outdoor spaces, enabling the RX the angle of arrival (AoA) and SNR for localization.

In this work, we focus on a single path LOS channel with a single measurement circumstance (we will discuss multiple measurements in future work).
We assume the RX applies a matched filter to the TX signal.  The resulting output is given by \cite{heath2018foundations}:
\begin{equation}
    \bs{z}[n] = g B(\phi-\phi^r)\bs{a}(\phi-\phi^r)\mathrm{sinc}((\tau-nT)/T) + \bs{w}[n],
\end{equation}
where, $\bs{z}[n] \in \C^{N_r}$ is the filtered response on each of the $N_r$ antennas at a sample $n$; $g$ is a complex channel gain; $\phi$ is the true AoA in the global coordinate system; $\phi^r$ is the orientation of the RX; $B(\cdot)$ is the element response
of the antenna; $\tau$ is time of arrival relative to the sampling window; and $T$
is the sampling period.
We assume that the angle of arrival is then estimated by maximizing the correlation 
\begin{equation} \label{eq:aoa_est}
    \wh{\phi} = \argmax_\phi \max_n
    |\bs{a}(\phi-\phi^r)\herm \bs{z}[n]|^2.
\end{equation}
The SNR is estimated from the value of the maximum:
\begin{equation} \label{eq:snr_est}
      \wh{\gamma} = \max_n
    |\bs{a}(\wh{\phi}-\phi^r)\herm \bs{z}[n]|^2.
\end{equation}
 We let $\bs{y}$ denote the observations:
\begin{equation} \label{eq:yobs}
    \bs{y} = (\bs{x}^r, \phi^r,
    \wh{\phi}, \wh{\gamma}),
\end{equation}
which includes both the receiver's position and orientation
 $(\bs{x}^r,\phi^r)$ as well as the
 the estimated angle of arrival $\wh{\phi}$ and estimated SNR $\wh{\gamma}$.
 Our basic problem is to estimate
 the posterior density $p(\bs{x}^t | \bs{y})$,
which represents the conditional distribution of the TX location
from the observations $\bs{y}$.



\subsection{Challenges}
\label{sec:challenges}
There are at least three complexities in the posterior from such angular estimates:

\noindent
\underline{\emph{False alarms}}: The estimate
\eqref{eq:aoa_est} is computed from the maximum of a number of angular hypotheses.
In low SNR settings, the strongest angular may correspond to noise, i.e., a false alarm.  The resulting posterior distribution will thus, in general, be a mixture of the distribution corresponding to a correct detection and the case where the peak arose from noise.  Each component of this distribution will have significantly different shapes.

\noindent
\underline{\emph{Angle ambiguity.}}  For uniform arrays, this array response  $\bs{a}(\phi)$ typically has angular symmetries. For example, in a uniform linear array, $\bs{a}(\phi)$ is a function of $\sin(\phi)$, so an angle $\phi$ and $\pi-\phi$ have identical responses.  This ambiguity creates a multi-modal distribution over two or more potential angles.

\noindent
\underline{\emph{Antenna Gain–strength coupling.}}
The antenna directional gain $B(\phi)$ of the RX enters the matched-filter statistics multiplicatively with the path gain $\alpha$.  In most antennas, particularly patch antennas, the antenna pattern is highly directional with significant front-to-back gain.  
As a result, the measured signal strength $|z|$ conflates two factors: true path loss and angular attenuation due to the beam pattern.  


\section{Proposed Solution}
\label{sec:method}
In this section, we introduce Monte Carlo Candidate Likelihood Estimation (MC-CLE), a data-driven framework for approximating the posterior distribution $p(\bs{x}^t|\bs{y})$ over a finite set of candidate transmit positions.


\subsection{Log Likelihood Estimation}
\label{sec:LLE}
We consider an estimate of the posterior probability distribution $p(\bs{x}^t|\bs{y})$, of the form:
\begin{equation}\label{eq:posterior_g}
    \wh{p}_\theta(\bs{x}^t|\bs{y}) = \frac{1}{Z_\theta(\bs{y})}
    e^{g_\theta(\bs{x}^t,\bs{y}) }p_0(\bs{x}^t),
\end{equation}
where $p_0(\bs{x}^t)$ is the known prior distribution on $\bs{x}^t$;
 $g_\theta(\cdot)$ is a learned
function of the position $\bs{x}^t$, observations $\bs{y}$
and parameters $\theta$; and $Z_\theta(\bs{y})$ is the partition function
\begin{equation} \label{eq:Zy}
    Z_\theta(\bs{y}) = \int 
    e^{g_\theta(\bs{x}^t,\bs{y})}p_0(\bs{x}^t)\, d\bs{x}^t.
\end{equation}
Under the distribution hypothesis 
\eqref{eq:posterior_g}, the corresponding estimate for the likelihood of $\bs{x}^t$
given $\bs{y}$ can be found from Bayes'
Rule:
\begin{align}
    \wh{p}(\bs{y}|\bs{x}^t) = \frac{
    \wh{p}(\bs{x}^t|\bs{y})p(y)}{Z_\theta(\bs{y}}
    = \frac{e^{g_\theta(\bs{x}^t,\bs{y})}
    }{Z_\theta(\bs{y})},
\end{align}
and $g_\theta(\bs{x}^t,\bs{y})$
can be considered
as an approximation of an un-normalized log likelihood:
\begin{equation} \label{eq:llapprox}
    g_\theta(\bs{x}^t,\bs{y}) \approx
    \log p(\bs{x}^t | \bs{y}) + \log \frac{Z_\theta(\bs{y})}{p(y)}.
\end{equation}
Hence, in the sequel, we will call $g_\theta(\cdot)$ the \emph{un-normalized log likelihood estimate}. As we will describe below, we parametrize $g_\theta(\cdot)$ as a neural network and learn its parameters $\theta$ directly from data.

Ideally, to train $g_\theta(\cdot)$, we would minimize a cost such as the cross-entropy \cite{CoverThomas06}:
\begin{align}
    H(p,\wh{p}_\theta) &:= -\Exp\left[ \log \wh{p}(\bs{x} |\bs{y}) \right] 
    \nonumber \\
    &= -\Exp[g_\theta(\bs{x}^t,\bs{y}) ]
    + H(p_0) + \Exp[ \log Z_\theta(\bs{y}) ],
    \label{eq:Hcross}
\end{align}
where the expectation is with respect to the true joint density $p(\bs{x}^t,\bs{y})$ and $H(p_0)=-\Exp \log p_0(\bs{x}^t)$ is the entropy of the prior.

\subsection{Sampled Cross Entropy Loss}
\label{sec:sampledloss}
The problem in the cross-entropy loss \eqref{eq:Hcross} is the expectation of the log partition function, $\Exp[ \log Z_\theta(\bs{y}) ]$. This normalization constant typically does not have 
an analytic expression when $g_\theta(\cdot)$ is a neural network.
To avoid this problem, we draw a \emph{candidate set}
$\mathcal{C}=\{\bar{\bs{x}}^t_k\}_{k=1}^K\subset\R^d$ and approximate the integral in \eqref{eq:Zy} by the Monte-Carlo sum 
\begin{equation} \label{eq:Zhat}
    \wh{Z}_\theta(\bs y)=\frac{1}{K}\sum_{k=1}^{K}
    e^{g_\theta(\bar{\bs{x}}^t_k,\bs y)}.
\end{equation} 
From the Law of Large Numbers, we have that
\begin{equation}
    \wh{Z}_\theta(\bs y) \rightarrow
    Z_\theta(\bs{y}).
\end{equation}
as $K \rightarrow \infty$ for any fixed 
$\bs{y}$.

Now suppose, that for training, we have data samples $(\bs{x}_i^t,\bs{y}_i)$, $i=1,\ldots,n$,
with true TX locations $\bs{x}^t_i$
along with measurements $\bs{y}_i$.
For each data sample $i$, we draw candidates $\ov{\bs{x}}_{ik}$, $k=1,\ldots,K$.
Replacing the expectation in \eqref{eq:Hcross}, and 
substituting the true log partition  $\log Z_\theta(\bs{y}_i)$ with the Monte-Carlo
estimate $\wh{Z}_\theta(\bs{y}_i)$, we obtain the \emph{sampled cross-entropy loss (CEL)} as:
\begin{align} \label{eq:sampled_CEL}
    \mc{L}(\theta) := \frac{1}{n}\sum_{i=1}^n
    \left[ -g_\theta(\bs{x}^t_i, \bs{y}_i) + \log\left[\frac{1}{K} \sum_{k=1}^K e^{g_\theta(\ov{\bs{x}}_{ik}^t,\bs{y}_i)} \right] \right].
\end{align}
We can then learn the parameters $\theta$
by minimizing this loss.

\subsection{Position Estimation with Neural Networks}
\label{sec:nn-estimator}

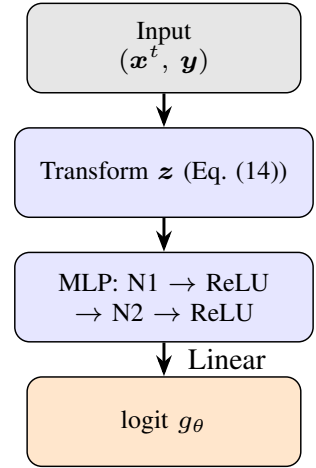
\begin{figure}[t]
  \centering
  \resizebox{.45\linewidth}{!}{%
  \begin{tikzpicture}[
    every node/.style={transform shape,font=\scriptsize},
    inb/.style  ={rectangle,draw=black,fill=gray!20,  rounded corners, minimum width=2.6cm,minimum height=0.85cm,align=center},
    blk/.style  ={rectangle,draw=black,fill=blue!10,  rounded corners, minimum width=2.6cm,minimum height=0.85cm,align=center,text width=2.6cm},
    outb/.style ={rectangle,draw=black,fill=orange!20,rounded corners, minimum width=2.6cm,minimum height=0.85cm,align=center},
    arr/.style  ={-{Stealth[length=2mm]},thick},
    node distance=3.2mm
  ]
    \node[inb] (in) {Input\\ $(\bs x^{t},\,\bs y)$};
    \node[blk, below=of in] (tf) {Transform $\bs z$ (Eq. \eqref{eq:zdef})};
    \node[blk, below=of tf] (mlp) {MLP: N1 $\to$ ReLU $\to$ N2 $\to$ ReLU};
    \node[outb, below=of mlp] (out) {logit $g_\theta$};

    \draw[arr] (in) -- (tf);
    \draw[arr] (tf) -- (mlp);
    \draw[arr] (mlp) -- node[right=1mm]{\footnotesize Linear} (out);
  \end{tikzpicture}%
  }
  \caption{Architecture of the MC-CLE model.}
  \label{fig:mccle_model}
\end{figure}

Having introduced the sampled cross-entropy loss
in \Cref{sec:sampledloss}, we now specify the function class
for $g_\theta(\bs{x}^t,\bs{y})$.

\textbf{Feature transformation:}
In the first step of the network, we apply a fixed transform to the input 
\[
    (\bs{x}^t,\bs{y}) = (\bs{x}^t,\bs{x}^r,\phi^r,\wh{\phi},\wh{\gamma}),
\]
to yield a feature vector $\bs{z}$ given by:
\begin{equation} \label{eq:zdef}
    \bs{z} = [\bs{u}, d, s, \cos(\phi^r),
    \sin(\phi^r), \cos(\wh{\phi}),
    \cos(\wh{\phi})],
\end{equation}
where $\bs{u}$ is a unit vector in the direction of $\bs{x}^t-\bs{x}^r$,
\begin{equation}
    \bs{u} = \frac{\bs{x}^t-\bs{x}^r}{\|\bs{x}^t-\bs{x}^r\|}.
\end{equation}
The feature $d$ is a scaled logarithmic distance:
\begin{equation} \label{eq:ddef}
    d = \log_{10}(\max\{1, \|\bs{x}^t-\bs{x}^r\|),
\end{equation}
where the distance $\|\bs{x}^t-\bs{x}^r\|$
is in meters;
and the feature $s$ is a scaled SNR in dB:
\begin{equation} \label{eq:sdef}
    s = \min \left\{ \max\left\{0, \frac{10}{s_{\rm max}}\log_{10}(\wh{\gamma}), 1 \right\}\right\},
\end{equation}
where $s_{\rm max} = $\,\SI{60}{dB}.
This particular transformation is physically-motivated, i.e., the likelihood should only depend on the relative position $\bs{x}^t-\bs{x}^r$.

In free-space propagation, the
distance effect is logarithmic \cite{heath2018foundations} -- hence the use of the logarithm in \eqref{eq:ddef}.
The clipping in both  \eqref{eq:ddef} and 
\eqref{eq:sdef} keep the variables in reasonable ranges for typical observed values.  The use of the sines and cosines
of the angles in \eqref{eq:zdef}
keeps the functions continuous modulo $2\pi$.


\textbf{Architecture:}
The transform feature vector $\bs{z}$
in \eqref{eq:zdef} has dimension $N_z=8$.
We then realize $g_\theta(\bs{x}^t,\bs{y})$ as two hidden linear layers of widths $N_1=64$ and~$N_2=16$,
each followed by \textsf{ReLU}. The last linear layer outputs a scalar -- see \Cref{fig:mccle_model}.

\textbf{Training objective:}
All parameters $\theta$ are trained end-to-end by minimizing the sampled cross-entropy loss 
\eqref{eq:sampled_CEL}.

\textbf{Gaussian Baselines:}
To demonstrate the utility of the method,
we compare against the case of a \emph{Gaussian posterior}.  Specifically, 
\begin{align}
    g_\theta(\bs{x}^t,\bs{y}) := -\frac{1}{2}
    (\bs{x}^t-\bs{\mu})^\intercal \bs{Q}
    (\bs{x}^t-\bs{\mu}),
\end{align}
where the mean $\mu \in \R^d$ and covariance
distribution $\bs{Q} \in \R^{d \times d}$
are neural network functions of the observations using the same structure except the output in \Cref{fig:mccle_model}.  We consider both the case where the Gaussian is in Cartesian and polar coordinates.

\begin{figure*}[t]
    \centering
    \includegraphics[width=0.95\linewidth]{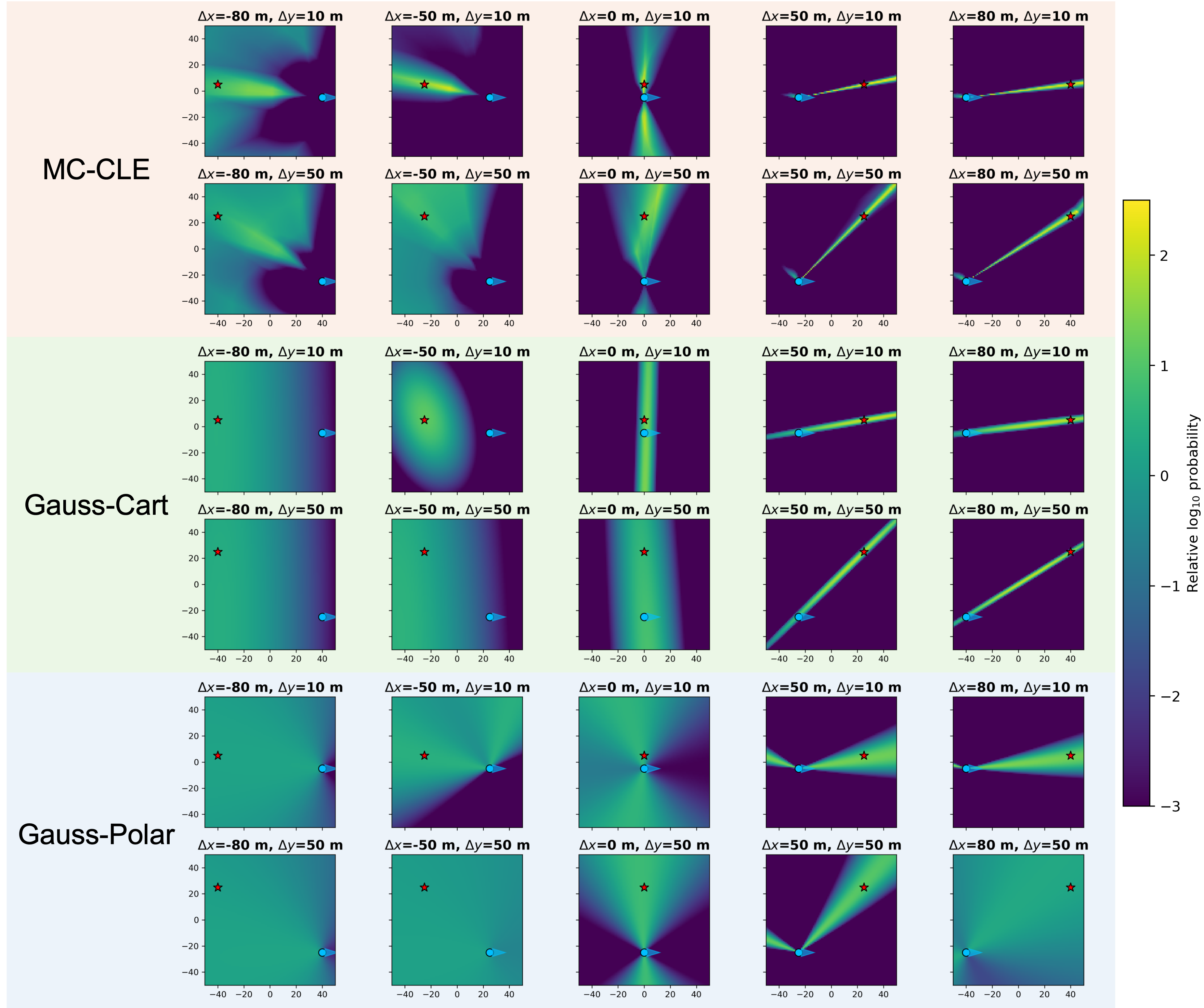}
    \caption{Relative log-probability maps for two RX–TX geometries (see \Cref{tab:rx-cases}). Columns sweep the horizontal offset $\Delta x\!\in\!\{-80,-50,0,50,80\}$\,m. Each panel spans a $100\times100$\,m region discretized on a dense grid. The red star marks the TX, the blue circle the RX, and the cyan arrow its heading. Colour encodes the relative log-probability $L(\bs{x}_k)=\log\!\big(p_\theta(\bs{x}_k,\bs{y})\,K_{\subsf grid}\big)$; warmer colours indicate larger $L(\bs{x}_k)$. Each horizontal strip corresponds to one model: MC–CLE (top), Gauss–Cart (middle), and Gauss–Polar (bottom). MC–CLE produces sharper ridges aligned with the TX–RX geometry; Gauss–Cart yields smoother, elongated bands; Gauss–Polar concentrates mass in narrower angular sectors with less accurate peak localization. When the RX faces the TX, high-confidence ridges appear; side- or back-looking orientations spread the posterior into broader, lower-confidence regions.}
    \label{fig:compare_heatmaps}
\end{figure*}

\section{Numerical Experiments}
\label{sec:experiments}
To evaluate MC-CLE, we use Sionna~\cite{sionna} as the simulation platform. All experiments operate at carrier frequency \SI{12}{GHz} and bandwidth of \SI{200}{MHz}.
This frequency is in the upper mid-band \cite{kang2024cellular}, which has attracted considerable interest for next generation cellular systems as well as localization \cite{raviv2024multi}.
The TX is modeled as a single isotropic antenna at a fixed, known height.
The RX uses the $8 \times 1$ azimuth-plane uniform linear array,
mounted at the same elevation as the TX to restrict the problem to two-dimensional localization.
We assume a 3GPP antenna pattern \cite{TR38.901} which has a \SI{3}{dB}
beamwidth of \SI{65}{\degree}.
Both TX and RX are generated uniformly in a $100 \times 100$\,\si{m} free-space environment, yielding a LOS scenario.
Since the distribution is uniform in the region, the prior $p_0(\bs{x}^t)$ is constant.
These settings are identical for all posterior models (MC-CLE, Gauss–Cart, and Gauss–Polar).

We adopt the network topology shown in \Cref{sec:nn-estimator}.
A total of $100{,}000$ TX–RX pairs are generated, and a $75\%/25\%$ train–validation split is used to select the best epoch during training.
All models use the same training configuration for $2{,}000$ epochs; MC-CLE uses AdamW with a learning rate $0.01$, and the Gaussian models use a learning rate $0.002$.
The MC-CLE training used $1.5$ GPU-hours, while each Gaussian model used $2.5$ GPU-hours, on NVIDIA RTX~A6000.

\subsection{CEL Performance}
Two evaluation sets of $25{,}000$ samples each are prepared:
\begin{itemize}
    \item \textbf{Grid-based set:} candidate positions are placed on a regular grid with approximately $3.33$\,m spacing,
    resulting in $961$ candidate positions within the $100 \times 100$\,m area.
    \item \textbf{Random set:} $1{,}000$ candidate positions are selected uniformly at random within the same area.
\end{itemize}


Given the sampled CEL $\mathcal{L}$ from \eqref{eq:sampled_CEL}
and the number of candidates $K=\{961,1000\}$ for the sets \{Grid-based, Random\}, we report the results in \Cref{tab:eval-losses}.
\begin{table}[t]
  \centering
  \caption{Evaluation losses for each model on grid and random evaluation sets.}
  \label{tab:eval-losses}
  \begin{tabular}{lrrr}
    \toprule
    Dataset   & MC-CLE   & Gauss–Cart & Gauss–Polar \\
    \midrule
    Eval Grid   & \num{-3.4703} & \num{-2.3441} & \num{-1.6788} \\
    Eval Random & \num{-3.4259} & \num{2.3071} & \num{-1.6634} \\
    \bottomrule
  \end{tabular}
\end{table}

To interpret the loss $\mathcal{L}$ in \eqref{eq:sampled_CEL}, first observe
that a uniform likelihood estimate, $g_\theta(\bs{x}^t,\bs{y})=$ constant, results in $\mc{L}=0$.  Also, recall that
the candidates $\ov{\bs{x}}^t_{ik}$ are selected such that the true position $\bs{x}_i^t$ is one of the candidates.
That is, $\bs{x}_i^t = \ov{\bs{x}}^t_{ik_0}$
for some $k_0 = \sigma(i)$.
Therefore, the sampled CEL in \eqref{eq:sampled_CEL} is bounded above by:
 \begin{align}
    \mc{L}(\theta) &\leq \frac{1}{n}\sum_{i=1}^n
    \left[ -g_\theta(\bs{x}^t_i, \bs{y}_i) + \log\left[\frac{1}{K}  e^{g_\theta(\bs{x}^t_i,\bs{y}_i)} \right] \right] \nonumber \\
    &= -\log(K), \label{eq:CEL_lower}
\end{align}
where the bound is achieved when the 
likelihood concentrates around the true location $\bs{x}^t=\bs{x}^t_i$. 
Hence, the loss is bounded by
\begin{equation}
    -\log(K) \leq \mc{L} \leq 0.
\end{equation}
This motivates two metrics:
\begin{align}
\mathcal{G} \triangleq e^{-\mathcal{L}}, \quad
\mathcal{R} \triangleq \frac{-\mathcal{L}}{\log K} \times 100\%.
\end{align}
Here $\mathcal{R}=0\%$ corresponds to uniform prediction ($\mathcal{L}_{\text{adj}}=0$), while
$\mathcal{R}=100\%$ corresponds to a perfect prediction $\mathcal{L}_{\text{adj}}=-\log K$.
The results are shown in \Cref{tab:eval-metrics}, and shows that the proposed method MC-CLE
significantly outperforms the Gaussian baselines.

\begin{table}[t]
  \centering
  \caption{Evaluation results reported as geometric improvement $\mathcal{G}$ and gap-closure ratio $\mathcal{R}$ (higher is better). $K{=}961$ for the grid set and $K{=}1000$ for the random set.}
  \label{tab:eval-metrics}
  \begin{tabular}{lcccccc}
    \toprule
    \multirow{2}{*}{Dataset} &
      \multicolumn{2}{c}{MC--CLE} &
      \multicolumn{2}{c}{Gauss--Cart} &
      \multicolumn{2}{c}{Gauss--Polar} \\
    \cmidrule(lr){2-3}\cmidrule(lr){4-5}\cmidrule(lr){6-7}
      & $\mathcal{G}$ & $\mathcal{R}$ & $\mathcal{G}$ & $\mathcal{R}$ & $\mathcal{G}$ & $\mathcal{R}$ \\
    \midrule
    Eval Grid & 32.15 & 50.53\% & 10.42 & 34.13\% & 5.36 & 24.44\% \\
    Eval Random & 30.75 & 49.59\% & 10.05 & 33.40\% & 5.28 & 24.08\% \\
    \bottomrule
  \end{tabular}
\end{table}

\subsection{Probability Mass Visualization}




\begin{table}[t]
  \caption{Five reused RX offset/heading cases ($\bs{x}^t$ is the
           fixed TX position).  Each row lists the RX offset
           $\bs{x}^r=(\Delta x,-\Delta y)$ and heading
           $\phi^{r}$, using $0^{\circ}\!=\!\text{right}(+x)$ and
           $90^{\circ}\!=\!\text{up}(+y)$.}
  \label{tab:rx-cases}
  \small
  \setlength{\tabcolsep}{4.5pt}
  \begin{tabularx}{\linewidth}{@{} c c c X @{}}
    \toprule
    \textbf{Row} & $\Delta y$ (m) & $\phi^{r}$ ($^{\circ}$) & Brief description\\
    \midrule
    1 & 10 & \phantom{0}0   & RX\,10\,m below TX, heading right\\
    2 & 50 & \phantom{0}0   & RX\,50\,m below TX, heading right\\
    \bottomrule
  \end{tabularx}
\end{table}

To visualize the ability of MC-CLE to capture the complexities of the posterior distribution, \Cref{fig:compare_heatmaps} plots 30 panels (3 models × 10 geometries) of the estimated probability distributions.  Each panel illustrates the estimated posterior distribution $\log p(\bs{x}^t,\bs{y})$ for one true TX location (red circle) and one RX location (blue circle) with the blue arrow representing the RX orientation.
The label for each plot $\Delta x, \Delta y$ represents the relative TX-RX position.
We observe that the learned distributions can capture significant features not possible under the assumption of a Gaussian posterior.  For example, when the TX is behind the RX ($\Delta x = -80$ and \SI{-50}{m}), the signal is weak due to the front-to-back gain of the antenna.  The MC-CLE captures the mixture distribution with one component in an angular cone corresponding to the estimated angle, and a second component
corresponding to the case where the TX location is not known, but somewhere behind the RX.  This mixture distribution cannot be captured in the Gaussian posteriors.  Additionally, when $\Delta x=50$ and \SI{80}{m}, the MC-CLE method is able to concentrate the distribution to a cone around the true angle $\phi$, but there is a second small cone for the angle $\pi-\phi$.

\begin{figure}[t]
    \centering
    \includegraphics[width=1.0\linewidth]{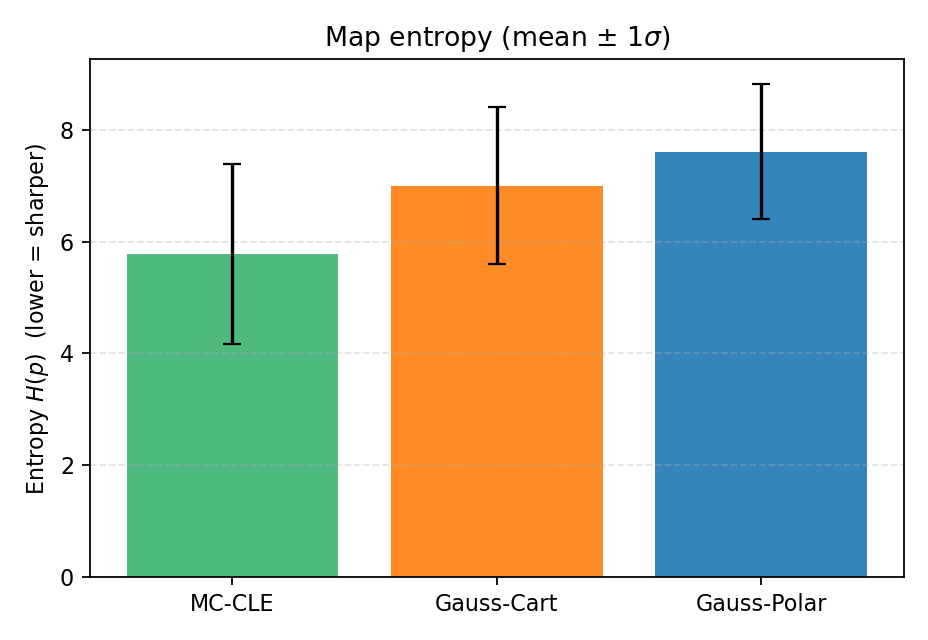}
    \caption{Posterior entropy across candidates, averaged over episodes for each model with error bars indicating $\pm 1\sigma$.
    Lower entropy indicates more concentrated posteriors.}
    \label{fig:posterior-entropy}
\end{figure}

Furthermore, we evaluate overall concentration via Shannon entropy in \Cref{fig:posterior-entropy}:
\begin{align}
    H(p) \triangleq -\sum_{i=1}^{K} p_i \log p_i .
\end{align}
Lower entropy indicates that probability mass is focused on fewer regions.
Empirically, MC-CLE yields lower entropy than the Gaussian baselines, indicating more concentrated posteriors.

\section{Conclusion}

This paper presents a simple method, \emph{Monte Carlo Candidate-Likelihood Estimation (MC-CLE)}, for estimation of the posterior distribution of the TX from RX measurements.  In a LOS setting, MC-CLE improves cross-entropy over Gaussian baselines and a uniform predictor, produces geometry-consistent ridges in relative log-probability maps, and yields lower posterior entropy. A calibrated spatial posterior provides a reliable bridge from signals to decisions, such as planning, control, and resource management.  Future work will consider how to extend these methods to NLOS settings, with and without
knowledge of the environment and obstacles.

\bibliographystyle{IEEEtran}
\bibliography{ref}

\end{document}